\documentclass{article}
\usepackage{amssymb}

\usepackage{graphicx}
\usepackage{amsmath}

\input{tcilatex}

\begin{document}

\begin{center}
{\Large Canonical Reduction of Symplectic Structures for the Maxwell and
Yang-Mills Equations. Part 1}

\bigskip

{\large A. Samoilenko}$^{(\ast )},${\large \ A. Prykarpatsky}$^{(\ast \ast
)} ${\large ,V. Samoylenko}$^{(\ast \ast \ast )}$

\bigskip

*) Institute of Mathematics at NAS, Kyiv 252601, Ukraina

**) Dept. of Applied Mathematics at the AGH, Krakow, 30059 Poland, and Dept.
of Physics at the EMU of\ Gazimagusa, N. Cyprus

(Emails: prykanat@cybergal.com, pryk.anat@excite.com)

*) Dept. of \ Mechanics and Mathematics at the National University, Kyiv
252601, Ukraina

(Email: vsam@imath.kiev.ua)

\bigskip
\end{center}

\bigskip \textit{Abstract}. The canonical reduction algorithm is applied to
Maxwell and Yang-Mills equations considered as Hamiltonian systems on some
fiber bundles with symplectic and connection structures. The minimum
interaction principle proved to have geometric origin within the reduction
method devised.

\textbf{0. Preliminaries}

We begin by reviewing the backgrounds of the reduction theory subject to
Hamiltonian systems \ with symmetry on principle fiber bundles. The material
is partly available in [1,4], so here will be only sketched but in notation
suitable for us.

Let $G$ denote a given Lie group with the unity element $e\in G$ and the
corresponding Lie algebra $\mathcal{G}$ $\simeq T_{e}(G).$ Consider a
principal fiber bundle $p:(M,\varphi )\rightarrow N$ with the structure
group $G$ and base manifold $N,$ on which the Lie group $G$ acts by means of
a mapping $\ \varphi :M\times G\rightarrow M.$ Namely, for each $g\in G$
there is a group diffeomorphism $\varphi _{g}:M\rightarrow M,$ generating
for any fixed $u\in M$ the following induced mapping: $\hat{u}:G\rightarrow
M $, where 
\begin{equation}
\hat{u}(g)=\varphi _{g}(u).  \tag{0.1}
\end{equation}

On the principal fiber bundle $p:(M,\varphi )\rightarrow N$ \ there is
assigned a connection $\Gamma ($\textit{$\mathcal{A}$}$)$ by means of such a
morphism \textit{$\mathcal{A}$}$\mathit{:}(T(M),\varphi _{g\ast
})\rightarrow (\mathcal{G},Ad),$ that for each $u\in M$ a mapping $\mathcal{A%
}(u):T_{u}(M)\rightarrow \mathcal{G}$ is a left inverse one to the mapping \ 
$\hat{u}_{\ast }(e):\mathcal{G}\rightarrow T_{u}(M),$ that is

\begin{equation}
\mathit{\mathcal{A}}(u)\hat{u}_{\ast }(e)=1.  \tag{0.2}
\end{equation}

Denote by $\Phi _{g}:T^{\ast }(M)\rightarrow T^{\ast }(M)$ the corresponding
lift of the mapping \ $\varphi _{g}:M\rightarrow M$ \ for all $g\in G.$ If $%
\alpha ^{(1)}\in \Lambda ^{1}(M)$ is the canonical $G$ - invariant 1-form on 
$\ M,$ a symplectic structure $\omega ^{(2)}\in \Lambda ^{2}(T^{\ast }(M))$
given by 
\begin{equation}
\omega ^{(2)}:=d\text{ }pr^{\ast }\alpha ^{(1)}  \tag{0.3}
\end{equation}
generates the corresponding momentum mapping $l:T^{\ast }(M)\rightarrow 
\mathcal{G}^{\ast },$ where 
\begin{equation}
l(\alpha ^{(1)})(u)=\hat{u}_{\ast }(e)\alpha ^{(1)}(u)  \tag{0.4}
\end{equation}
for all $u\in M.$ Remark here that the principal fiber \ bundle structure $%
p:(M,\varphi )\rightarrow N$ \ means in part the exactness of the following
sequences of mappings: 
\begin{equation}
0\rightarrow \mathcal{G}\overset{\hat{u}_{\ast }(e)}{\rightarrow }T_{u}(M)%
\overset{p_{\ast }(u)}{\rightarrow }T_{p(u)}(N)=0,  \tag{0.5}
\end{equation}
that is 
\begin{equation}
p_{\ast }(u)\hat{u}_{\ast }(e)=0  \tag{0.6}
\end{equation}
for all $u\in M.$ Combining (0.6) with (0.2) and (0.4), one obtains such an
embedding: 
\begin{equation}
\lbrack 1-\mathcal{A}^{\ast }(u)\hat{u}^{\ast }(e)]\alpha ^{(1)}(u)\in range%
\text{ }p^{\ast }(u)  \tag{0.7}
\end{equation}
for each canonical 1-form $\alpha ^{(1)}\in \Lambda ^{1}(M)$ at $u\in M.$
The expression (0.7) means of course, that 
\begin{equation}
\hat{u}^{\ast }(e)[1-\mathcal{A}^{\ast }(u)\hat{u}^{\ast }(e)]\alpha
^{(1)}(u)=0  \tag{0.8}
\end{equation}
for all $u\in M.$ Taking now into account that the mapping \ $p^{\ast
}(u):T^{\ast }(N)\rightarrow T^{\ast }(M)$ \ is for each $u\in M$ injective,
it has the unique inverse mapping \ $\ (p^{\ast }(u))^{-1}$ upon its image \ 
$p^{\ast }(u)T^{\ast }(N)\subset T^{\ast }(M).$ Thereby \ for each $u\in M$
one can define a morphism $p_{\mathcal{A}}:(T^{\ast }(M),\Phi )\rightarrow
T^{\ast }(N)$ as 
\begin{equation}
p_{\mathcal{A}}(u):\alpha ^{(1)}\rightarrow (p^{\ast }(u))^{-1}[1-\mathcal{A}%
^{\ast }(u)\hat{u}^{\ast }(e)]\alpha ^{(1)}(u).  \tag{0.9}
\end{equation}

Based on the definition (0.9) one can easily check that the diagram 
\begin{equation}
\begin{array}{ccc}
T^{\ast }(M) & \overset{p_{\mathcal{A}}}{\rightarrow } & T^{\ast }(N) \\ 
\left. pr\right\downarrow &  & \left\downarrow pr\right. \\ 
M & \overset{p}{\rightarrow } & N
\end{array}
\tag{0.10}
\end{equation}
is commutative.

Let \ now an element $\xi \in \mathcal{G}^{\ast }$ be $G-$invariant, that is 
$\ \ Ad_{g}^{\ast }\xi =\xi $ for all $\ g\in G.$ Denote also by \ $p_{%
\mathcal{A}}^{\xi }$ \ the restriction of the mapping (0.9) upon the subset $%
l^{-1}(\xi )\in T^{\ast }(M),$ that is \ \ $p_{\mathcal{A}}^{\xi
}:l^{-1}(\xi )\rightarrow T^{\ast }(N)$, where for all $u\in M$%
\begin{equation}
p_{\mathcal{A}}^{\xi }(u):l^{-1}(\xi )\rightarrow (p^{\ast }(u))^{-1}[1-%
\mathcal{A}^{\ast }(u)\hat{u}^{\ast }(e)]l^{-1}(\xi ).  \tag{0.11}
\end{equation}
Now one can characterize the structure of the reduced phase space $%
l^{-1}(\xi )/G$ \ by means of the following theorem.

\textbf{Lemma 0.1}{\large \ }\textit{The mapping }$p_{\mathcal{A}}^{\xi
}(u):l^{-1}(\xi )\rightarrow T^{\ast }(N)$\textit{\ is a principal fiber }$G$%
\textit{-bundle with the reduced space \ }$l^{-1}(\xi )/G$\textit{\ being
diffeomorphic to }$T^{\ast }(N).$

Denote by $<\cdot ,\cdot >_{\mathcal{G}}$ the standard $Ad$-invariant
nondegenerate scalar product on $\mathcal{G}^{\ast }\times \mathcal{G}.$\ \
Based on Lemma 0.1 one derives the following \ characteristic theorem.

\textbf{Theorem 0.2} \ \textit{Given a principal fiber \ }$G$\textit{-bundle
with a connection }$\Gamma (\mathcal{A})$\textit{\ and a }$G$\textit{%
-invariant element }$\ \xi \in \mathcal{G}^{\ast },$\textit{\ then each such
a connection }$\Gamma (A)$\textit{\ defines a symplectomorphism }$\nu _{\xi
}:l^{-1}(\xi )/G\rightarrow T^{\ast }(N)$\textit{\ between the reduced phase
space }$l^{-1}(\xi )/G$\textit{\ and \ cotangent bundle \ }$T^{\ast }(N),$%
\textit{\ where }$l:T^{\ast }(M)\rightarrow \mathcal{G}^{\ast }$\textit{\ is
the naturally associated momentum mapping for the group }$G$\textit{-action
on }$M.$\textit{\ Moreover, the following equality } 
\begin{equation}
(p_{\mathcal{A}}^{\xi })(d\text{ }pr^{\ast }\beta ^{(1)}+pr^{\ast }\text{ }%
\Omega _{\xi }^{(2)})=\left. d\text{ }pr^{\ast }\alpha ^{(1)}\right|
_{l^{-1}(\xi )}  \tag{0.12}
\end{equation}
\textit{holds for the canonical 1-forms \ }$\beta ^{(1)}\in \Lambda ^{1}(N)$%
\textit{\ and \ }$\alpha ^{(1)}\in \Lambda ^{1}(M),$\textit{\ where \ }$%
\Omega _{\xi }^{(2)}:=<\Omega ^{(2)},\xi >_{\mathcal{G}}$\textit{\ is the }$%
\xi $\textit{-component of the corresponding curvature form }$\Omega
^{(2)}\in \Lambda ^{(2)}(N)\otimes \mathcal{G}.$

\textbf{Remark 0.3} \textit{As the canonical 2-form \ \ \ }$d$\textit{\ }$%
pr^{\ast }\alpha ^{(1)}\in $\textit{\ }$\Lambda ^{(2)}(T^{\ast }(M))$\textit{%
\ \ \ is }$G$\textit{-invariant on \ \ \ }$T^{\ast }(M)$\textit{\ due to
construction, it is evident that its restriction upon the \ }$G$\textit{%
-invariant submanifold \ \ \ }$l^{-1}(\xi )\subset T^{\ast }(M)$\textit{\ \
will be effectively defined only on the reduced space \ \ }$l^{-1}(\xi )/G,$%
\textit{\ that ensures the validity of the equality sign in (0.12). \ }

As a consequence of Theorem 0.2 one can formulate the following useful
enough for applications results.

\textbf{Theorem 0.4} \textit{Let an element }$\xi \in \mathcal{G}^{\ast }$%
\textit{\ has the isotropy group }$G_{\xi }$\textit{\ \ \ acting on the
subset \ \ }$l^{-1}(\xi )\subset T^{\ast }(M)$\textit{\ freely and properly,
so that the reduced phase space \ }$(l^{-1}(\xi )/G\ ,\sigma _{\xi }^{(2)})$%
\textit{\ is symplectic, where by definition, } 
\begin{equation}
\sigma _{\xi }^{(2)}:=\left. d\text{ }pr^{\ast }\alpha ^{(1)}\right|
_{l^{-1}(\xi )}.  \tag{0.13}
\end{equation}
\textit{If a principal fiber bundle \ \ }$p:(M,\varphi )\rightarrow N$%
\textit{\ has a structure group coinciding with }$G_{\xi }$\textit{\ , then\
\ the reduced symplectic space \ \ \ \ }$(l^{-1}(\xi )/G_{\xi },\sigma _{\xi
}^{(2)})$\textit{\ is symplectomorphic to the cotangent symplectic space }$%
(T^{\ast }(N),\omega _{\xi }^{(2)})$\textit{, where } 
\begin{equation}
\omega _{\xi }^{(2)}=d\text{ }pr^{\ast }\beta ^{(1)}+pr^{\ast }\Omega _{\xi
}^{(2)},  \tag{0.14}
\end{equation}
\textit{and the corresponding symplectomorphism \ is given by the relation
like (0.12).}

\textbf{Theorem 0.5} \textit{In order that two symplectic spaces }$%
(l^{-1}(\xi )/G_{\xi },\sigma _{\xi }^{(2)})$\textit{\ and \ }$(T^{\ast
}(N),d$\textit{\ }$pr^{\ast }\beta ^{(1)})$ were \textit{symplectomorphic,
it is necessary and sufficient that the element \ }$\xi \in \ker $\textit{\
h, where for G-invariant element \ \ }$\xi \in \mathcal{G}^{\ast }$\textit{\
the mapping }$h:\xi \rightarrow \lbrack \Omega _{\xi }^{(2)}]\in H^{2}(N;%
\mathbb{Z})$\textit{\ with }$H^{2}(N;\mathbb{Z})$\textit{\ being the
cohomology class of 2-forms \ on the manifold }$N.$

In case when there is given a Lie group $G,$\ the tangent space \ $T(G)$\ is
also a Lie group isomorphic to the semidirect product\textit{\ }$\tilde{G}%
:=G\circledast _{Ad}\mathcal{G}$ of the Lie group $G$ and its Lie algebra $%
\mathcal{G}$ under the adjoint action $Ad$ of $G$ on $\mathcal{G}.$ The Lie
algebra $\mathcal{\tilde{G}}$ of \ $\tilde{G}$ is correspondingly the
semidirect product of $\mathcal{G}$ with itself , regarded as a trivial
abelian Lie algebra, under the adjoint action $ad$ and has thus the bracket
\ defined by $%
[(a_{1},m_{1}),(a_{2},m_{2})]:=([a_{1},a_{2}],[a_{1},m_{2}]+[a_{2},m_{1}])$
for all $(a_{j},m_{j})\in \mathcal{G}\circledast _{ad}\mathcal{G},$\ $j=%
\overline{1,2}.$ Take now any element \ $\xi \in \mathcal{G}^{\ast }$ and
compute its isotropy group $G_{\xi }$ under the coadjoint action \ $Ad^{\ast
}$ \ of $\ G$ on \ $\mathcal{G}^{\ast },$ and denote by \ $\mathcal{G}_{\xi
} $ its Lie algebra. The cotangent bundle $T^{\ast }(G)$ is obviously
diffeomorphic to $M:=G\times \mathcal{G}^{\ast \text{ \ \ }}$on which the
Lie group $G_{\xi }$ acts freely and properly (due to construction) by left
translation on the first factor and Ad$^{\ast }$ -action on the second one.
The corresponding momentum mapping $\ l:G\times \mathcal{G}^{\ast
}\rightarrow \mathcal{G}_{\xi }^{\ast }$ is obtained as 
\begin{equation}
l(h,\alpha )=\left. Ad_{h^{-1}}^{\ast }\alpha \right| _{\mathcal{G}_{\xi
}^{\ast }}  \tag{0.15}
\end{equation}
with no critical point. Let now $\eta \in \mathcal{G}^{\ast }$ and $\eta
(\xi ):=\left. \eta \right| _{\mathcal{G}_{\xi }^{\ast }}.$ Therefore the
reduced space \ $(l^{-1}(\eta (\xi ))/G_{\xi }^{\eta (\xi )},\sigma _{\xi
}^{(2)})$ has to be symplectic due to the well known Marsden-Weinstein
reduction theorem [2,5], where $G_{\xi }^{\eta (\xi )}$ \ is the isotropy
subgroup of the $G_{\xi }$ -coadjoint action on \ $\eta (\xi )\in \mathcal{G}%
_{\xi }^{\ast }$ and the symplectic form \ $\sigma _{\xi }^{(2)}:=\left. d%
\text{ }pr^{\ast }\alpha ^{(1)}\right| _{l^{-1}(\eta (\xi ))}$ is naturally
induced \ from the canonical symplectic structure on $T^{\ast }(G).$ Define
now for $\eta (\xi )\in \mathcal{G}_{\xi }^{\ast }$ the one-form $\alpha
_{\eta (\xi )}^{(1)}\in \Lambda ^{1}(G)$ as 
\begin{equation}
\alpha _{\eta (\xi )}^{(1)}(h):=R_{h}^{\ast }\eta (\xi ),  \tag{0.16}
\end{equation}
where $R_{h}:G\rightarrow G$ is right translation by an element $h\in G.$ It
is easy to check that the element (0.16) is right $G$-invariant and left $%
G_{\xi }^{\eta (\xi )}$-invariant, thus inducing a one-form on the quotient $%
N_{\xi }:=G/G_{\xi }^{\eta (\xi )}.$ Denote by $pr^{\ast }\alpha _{\eta (\xi
)}^{(1)}$ its pull-back to $T^{\ast }(N_{\xi })$ \ \ and form the symplectic
manifold $(T^{\ast }(N_{\xi }),d$ $pr^{\ast }\beta ^{(1)}+d$ $pr^{\ast
}\alpha _{\eta (\xi )}^{(1)}),$ where \ $d$ $pr^{\ast }\alpha _{\eta (\xi
)}^{(1)}$\ \ $\in \Lambda ^{(2)}(T^{\ast }(N_{\xi }))$\ \ is the canonical
symplectic form on \ \ $T^{\ast }(N_{\xi }).$ The construction above now can
be summarized as the next theorem.

\textbf{Theorem 0.6}\ \textit{Let }$\xi ,\eta \in \mathcal{G}^{\ast }$ and \ 
$\eta (\xi ):=\left. \eta \right| _{\mathcal{G}_{\xi }^{\ast }}$ \ \textit{%
be fixed}. \textit{Then the reduced symplectic manifold }$(l^{-1}(\eta (\xi
))/G_{\xi }^{\eta (\xi )},\sigma _{\xi }^{(2)})$\textit{\ is a symplectic
covering of the coadjoint orbit }$Or(\xi ,\eta (\xi );\tilde{G})$\textit{\
and symplectically embeds onto a subbundle over }$\ G/G_{\xi }^{\eta (\xi )}$%
\textit{\ of\ \ }$(T^{\ast }(G/G_{\xi }^{\eta (\xi )}),$\textit{\ }$\omega
_{\xi }^{(2)}),$\textit{\ with }$\omega _{\xi }^{(2)}:=d$\textit{\ }$%
pr^{\ast }\beta ^{(1)}+d$\textit{\ }$pr^{\ast }\alpha _{\eta (\xi
)}^{(1)}\in \Lambda ^{2}(T^{\ast }(G/G_{\xi }^{\eta (\xi )}).$

The statement above fits into the conditions of \ Theorem 0.4 if one to
define a connection 1-form $\mathcal{A}(g):T_{g}(G)\rightarrow \mathcal{G}%
_{\xi }$\bigskip\ \ as follows: 
\begin{equation}
<\mathcal{A}(g),\xi >_{\mathcal{G}}:=R_{g}^{\ast }\eta (\xi )  \tag{0.17}
\end{equation}
for any $\xi \in \mathcal{G}^{\ast }.$ \ The expression (0.17) generates a
completely horizontal \ 2-form \ $d<\mathcal{A}(g),\xi >_{\mathcal{G}}$ on
the Lie group $G,$ which gives rise immediately to the symplectic structure $%
\omega _{\xi }^{(2)}$ on the reduced phase space $T^{\ast }(G/G_{\xi }^{\eta
(\xi )}).$

\bigskip

\textbf{1 The Maxwell electromagnetic equations .}

\bigskip Under the Maxwell electromagnetic equations we shall understand the
following relationships on a cotangent phase space $T^{\ast }(N)$ \ with $%
N\subset \mathit{T}(D;\mathbb{R}^{3})$ \ being a manifold of vector fields
on some almost everywhere smooth enough domain $D\subset \mathbb{R}^{3}$:

\begin{eqnarray}
\partial E/\partial t &=&rotB,\text{ \ \ \ }\partial B/\partial t=-rotE 
\TCItag{1.1} \\
divE &=&\rho ,\text{ \ \ \ \ \ \ \ \ }divB=0,  \notag
\end{eqnarray}
where $(E,B)\in T^{\ast }(N)$ \ \ is a vector of electric and magnetic
fields and $\rho \in C(D;\mathbb{R})$ \ is some fixed density function for a
smeared out ambient charge.

Aiming to represent equations (1.1) as those on reduced symplectic space,
define as \ in [9] the appropriate configuration space $M$ $\subset \mathcal{%
T}(D;\mathbb{R}^{3}),$ with a vector potential field coordinate \ $A\in M.$
The cotangent space $T^{\ast }(M)$\ \ may be identified with pairs $(\
A,Y)\in T^{\ast }(M)$, where $Y\in \mathcal{T}^{\ast }(D;\mathbb{R}^{3})$ is
a vector field density in $D.$ On the space $T^{\ast }(M)$ there exists the
canonical symplectic form $\omega ^{(2)}\in \Lambda ^{2}(T^{\ast }(M),$ where%
$\ \ \ \omega ^{(2)}:=dpr^{\ast }\alpha ^{(1)},\ $and $\ \ \ \ \ \ \ \ \ \ \
\ \ \ \ \ \ $%
\begin{equation}
\alpha ^{(1)}(A,Y)=\int_{D}d^{3}x<Y,dA>:=(Y,dA),  \tag{1.2}
\end{equation}
where by \ $<\cdot ,\cdot >$ we denoted the standard scalar product in \ $%
\mathbb{R}^{3}$ endowed with the measure $d^{3}x$, and by $pr:T^{\ast
}(M)\rightarrow M$ we denoted the usual basepoint \ projection upon the base
space $M.$ Define now a Hamiltonian function $H\in \mathcal{D}(T^{\ast }(M))$
as 
\begin{equation}
H(A,Y)=1/2((Y,Y)+(rotA,rotA)),  \tag{1.3}
\end{equation}
which is evidently invariant with respect to the following symmetry group $G$
acting on the base manifold $M$ and lifted to $T^{\ast }(M):$ for any $\psi
\in \mathcal{G}\subset C^{(1)}(D;\mathbb{R})$ and $(A,Y)\in T^{\ast }(M)$%
\begin{equation}
\varphi _{\psi }(A):=A+\nabla \psi ,\text{ \ \ \ \ \ \ }\Phi _{\psi }(Y)=Y. 
\tag{1.4}
\end{equation}
Under the transformation (1.4) the 1-form (1.2) is evidently invariant too
since 
\begin{equation}
\varphi _{\psi }^{\ast }\alpha ^{(1)}(A,Y)=(Y,dA+\nabla d\psi )=-(divY,d\psi
)=\alpha ^{(1)}(A,Y),  \tag{1.5}
\end{equation}
where we made use of the condition that $d\psi \simeq 0$ in $\Lambda
^{1}(M). $ Thus, the corresponding momentum mapping (0.4) is given as

\bigskip

\begin{equation}
l(A,Y)=-divY  \tag{1.6}
\end{equation}
for all $(A,Y)\in T^{\ast }(M).$ If $\rho \in \mathcal{G}^{\ast },$ where $%
\mathcal{G}$ \ is the corresponding to $G$ \ Lie algebra, one can define the
reduced space $l^{-1}(\rho )/G,$ since evidently, the isotropy group $%
G_{\rho }=G$ due to its commutativity. Consider \ \ now a principal fiber
bundle $p:M\rightarrow N$ with the abelian structure group $G$ and a base
manifold $N$ taken as 
\begin{equation}
N:=\{B\in \mathcal{T}(D;\mathbb{R}^{3}):divB=0\},  \tag{1.7}
\end{equation}
where, by definition 
\begin{equation}
p(A):=B=rotA.  \tag{1.8}
\end{equation}
Over this bundle one can build a connection 1-form $\ \ \mathcal{A}%
:T(M)\rightarrow \mathcal{G}\mathbf{,}$ where for all $A\in M$ 
\begin{equation}
\mathcal{A}(A)\cdot \hat{A}_{\ast }(l)=1,\text{ \ \ }d<\mathcal{A}(A),\rho
>_{\mathcal{G}}=\Omega _{\rho }^{(2)}(B)  \tag{1.9}
\end{equation}
in virtue of commutativity of the Lie algebra $\mathcal{G}$. Then, due to
Theorem 0.4 the cotangent manifold $T^{\ast }(N)$ is symplectomorphic to the
reduced phase space $l^{-1}(\rho )/G\cong \{(B,E)\in T^{\ast }(N):divE=\rho
, $ $divB=0\}$ with the canonical symplectic 2-form 
\begin{equation}
\omega _{\rho }^{(2)}(B,E)=(dS,\wedge dB)+d<\mathcal{A}(A),\rho >_{\mathcal{G%
}},  \tag{1.10}
\end{equation}
where we put $\ rot$ $S=-E.$ The Hamiltonian (1.3) reduces correspondingly
to the following classical form:

\begin{equation}
H(B,E)=1/2((B,B)+(E,E)).  \tag{1.11}
\end{equation}
As a result, the Maxwell equations (1.1) become a Hamiltonian system upon
the reduced phase space $T^{\ast }(N)$ endowed with the quasicanonical
symplectic structure (1.10) and \ the new Hamiltonian function (1.11).

It is well known that Maxwell equations (1.1) admit a one more canonical
symplectic structure on $T^{\ast }(N),$ namely 
\begin{equation}
\bar{\omega}^{(2)}:=(dB,\wedge dE),  \tag{1.12}
\end{equation}
with respect to which they are Hamiltonian too and whose ''helicity''
conservative Hamiltonian function reads as 
\begin{equation}
\bar{H}(B,E)=1/2((rotE,E)+(rotB,B)),  \tag{1.13}
\end{equation}
where $(B,E)\in T^{\ast }(N).$ It easy to see that (1.13) is also an
invariant function \ with respect to the Maxwell equations (1.1). Subject to
the Maxwell equations (1.1) a group theoretical interpretation of the
symplectic structure (1.12) is still waiting for search.

Notice now that both symplectic structure (1.12) and Hamiltonian (1.13) are
invariant with respect to the following abelian group $\ G^{2}=G\times G$\
-action: 
\begin{equation}
G^{2}\ni (\psi ,\chi ):(B,E)\rightarrow (B+\nabla \psi ,E+\nabla \chi ) 
\tag{1.14}
\end{equation}
for all $(B,E)\in T^{\ast }(N).$ \ Corresponding to (1.14) the momentum
mapping \ $l:T^{\ast }(N)\rightarrow \mathcal{G}^{\ast }\mathbf{\times }%
\mathcal{G}^{\ast }$ \ is calculated out as $\ \ \ \ $%
\begin{equation}
\ l(B,E)=(divE,-divB)  \tag{1.15}
\end{equation}
for any $(B,E)\in T^{\ast }(N).$ Fixing a value of (1.15) as $l(B,E)=\xi
:=(\rho ,0),$ that is 
\begin{equation}
divE=\rho ,\text{ \ }divB=0,  \tag{1.16}
\end{equation}
one obtains the reduced phase space $l^{-1}(\xi )/G^{2},$ since the isotropy
subgroup $G_{\xi }^{2}$ of the element $\xi \in \mathcal{G}^{\ast }\mathbf{%
\times }\mathcal{G}^{\ast }$ coincides with entire group $G^{2}.$ Thus the
reduced phase space due to Theorem 0.4 is endowed with the canonical
symplectic structure 
\begin{equation}
\bar{\omega}^{(2)}(A,Y)=(dY,\wedge dA)+d<\mathcal{A}(A),\xi >_{\mathcal{G}},
\tag{1.17}
\end{equation}
where $T^{\ast }(M)\ni (A,Y)$ are variables constituting the corresponding
coordinates upon the cotangent space over an associated fibre bundle $\bar{p}%
:N\rightarrow M$ \ with a curvature 1-form $\mathcal{A}:T(N)\rightarrow 
\mathcal{G}\mathbf{\times }\mathcal{G}\mathbf{.}$ In virtue of (1.16) one
can define the projection map $\bar{p}:N\rightarrow M$ as follows: 
\begin{equation}
\bar{p}(B):=rot^{-1}B=A  \tag{1.18}
\end{equation}
for any $A\in M\in \mathcal{T}(D;\mathbb{R}^{3}).$ It is evident that the
second condition of (1.16) is satisfied automatically upon the cotangent
bundle $T^{\ast }(M).$ Subject to the coadjoint variables $Y\in T_{A}^{\ast
}(M)$ and $E\in T_{B}^{\ast }(N)$ \ \ for all $A\in M$ and $E\in N$ one can
easily obtain from the equality \ $\bar{p}^{\ast }\beta ^{(1)}=\alpha ^{(1)}$
the expression 
\begin{equation}
Y=-rotE,  \tag{1.19}
\end{equation}
satisfying the evident condition $divY=0.$ The Hamiltonians (1.11) and
(1.13) take correspondingly on $T^{\ast }(M)$ the forms as 
\begin{equation}
\mathcal{\bar{H}}(A,Y)=1/2((rot^{3}A,A)+(rot^{-1}Y,Y)),  \tag{1.20}
\end{equation}
and

\begin{equation*}
\mathcal{H}(A,Y)=1/2((rot^{-1}Y,rot^{-1}Y)+(rotA,rotA)),
\end{equation*}
being obviously invariant too with respect to common evolutions on $T^{\ast
}(M).$ As was mentioned in [1], the invariant like (1.13) admits the
following geometrical interpretation: its quantity is a related with
dynamical equations helicity structure, that is a number of closed linkages
of the vortex lines present in the ambient phase space.

If one to consider now a motion of a charged particle under a Maxwell field,
it is convenient to introduce another fiber bundle structure $p:M\rightarrow
N,$ \ namely such one that $M=N\times G$\bigskip , $\ N:=D\subset \mathbb{R}%
^{3}$ and $G:=\mathbb{R}/\{0\}$ being the corresponding (abelian) structure
Lie group. An analysis similar to the above gives rise to a reduced upon the
space $l^{-1}(\xi )/G\simeq T^{\ast }(N),$ $\xi \in \mathcal{G}$, symplectic
structure $\omega ^{(2)}(q)=<dp,\wedge dq>+d<\mathcal{A}(q,g),\xi >_{%
\mathcal{G}},$ where $\mathcal{A}(q,g):=<A(q),dq>+g^{-1}dg$ is a usual
connection 1-form on $\ \ M,$ with $(q,p)\in T^{\ast }(N)$ and $g\in G.$ The
corresponding canonical Poisson brackets on $T^{\ast }(N)$ are easily found
to be 
\begin{equation}
\{q^{i},q^{j}\}=0,\text{ \ \ }\{p_{j},q^{i}\}=\delta _{j}^{i},\text{ \ \ \ \
\ \ }\{p_{i},p_{j}\}=F_{ji}(q)  \tag{1.21}
\end{equation}
for all $(q,p)\in T^{\ast }(N).$ If one introduces a new momentum variable $%
\tilde{p}:=p+A(q)$ on $T^{\ast }(N)\ni (q,p),$ it is easy to verify that \ $%
\omega _{\xi }^{(2)}\rightarrow \tilde{\omega}_{\xi }^{(2)}:=<d\tilde{p}%
,\wedge dq>$, giving rise to the following Poisson brackets [8]: 
\begin{equation}
\{q^{i},q^{j}\}=0,\text{ \ \ \ \ }\{\tilde{p}_{j},q^{i}\}=\delta _{j}^{i},%
\text{ \ \ \ \ \ \ }\{\tilde{p}_{i},\tilde{p}_{j}\}=0,  \tag{1.22}
\end{equation}
where $i,j=\overline{1,3},$ iff for all $i,j,k=\overline{1,3}$ \ the
standard Maxwell field equations are satisfied on $N:$%
\begin{equation}
\partial F_{ij}/\partial q_{k}+\partial F_{jk}/\partial q_{i}+\partial
F_{ki}/\partial q_{j}=0  \tag{1.23}
\end{equation}
with the carvature tensor $F_{ij}(q):=\partial A_{j}/\partial q^{i}-\partial
A_{i}/\partial q^{j},$ \ $i,j=\overline{1,3},$ $q\in N.$ Such a construction
permits a natural generalization to the case of nonabelian structure Lie
group yielding a description of Yang-Mills field equations within the
reduction approach.

\bigskip

\textbf{2. A charged particle phase space structure and Yang-Mills field
equations. }

As before, we start with defining a phase space $M$ of a particle \ \ \
under a Yang-Mills field in a region $D\subset \mathbb{R}^{3}$ as $M:=D%
\mathbb{\times }G,$ where $G$ is a (not in general semisimple) Lie group,
acting on $M$ \ from the right. Over the space $M$ one can define quite
naturally a connection $\Gamma (\mathcal{A})$ \ if to consider the following
trivial principal fiber bundle $p:M\rightarrow N,$ where $N:=D,$ with the
structure group $G.$ Namely, if $g\in G$, $\ q\in N,$ then a connection
1-form on $M\ni (q,g)$ can be written down [1,3,7] as 
\begin{equation}
\mathcal{A}(q;g):=g^{-1}(d+\sum_{i=1}^{n}a_{i}A^{(i)}(q))g,  \tag{2.1}
\end{equation}
where $\{a_{i}\in \mathcal{G}:i=\overset{}{\overline{1,n}}\}$ is a basis of
the Lie algebra $\mathcal{G}$ \ of the Lie group $G$, and $%
A_{i}:D\rightarrow \Lambda ^{1}(D),$ $i=\overline{1,n},$ are the Yang-Mills
fields in the physical space \ \ $D\subset \mathbb{R}^{3}.$

Now one defines the natural left invariant Liouville form on $M$ \ as 
\begin{equation}
\alpha ^{(1)}(q;g):=<p,dq>+<y,g^{-1}dg>_{\mathcal{G}},  \tag{2.2}
\end{equation}
where $y\in T^{\ast }(G)$ and $\ <\cdot ,\cdot >_{\mathcal{G}}$ denotes \ as
before the usual Ad-invariant nondegenerate bilinear form on $\mathcal{G}%
^{\ast }\times \mathcal{G},$ as evidently $g^{-1}dg\in \Lambda
^{1}(G)\otimes \mathcal{G}\mathbf{.}$ The main assumption we need to accept
for further is that the connection 1-form is in accordance with the Lie
group $G$ \ action on $M.$ The latter means that the condition 
\begin{equation}
R_{h}^{\ast }\mathcal{A}(q;g)=Ad_{h-1}\mathcal{A}(q;g)  \tag{2.3}
\end{equation}
is satisfied for all \ \ $(q,g)\in M$ and $h\in G,$ where $%
R_{h}:G\rightarrow G$ means the right translation by an element $h\in G$ on
the Lie group $\ G.$

Having stated all preliminary conditions needed for the reduction Theorem
0.4 to be applied to our model, suppose that the Lie group $G$ canonical
action on $M$ is naturally lifted to that on the cotangent space $T^{\ast
}(M)$ endowed due to (2.2) with the following $G$-invariant canonical
symplectic structure: \ \ \ \ 
\begin{eqnarray}
\omega ^{(2)}(q,p;g,y) &:&=d\text{ }pr^{\ast }\alpha
^{(1)}(q,p;g,y)=<dp,\wedge dq>  \TCItag{2.4} \\
+ &<&dy,\wedge g^{-1}dg>_{\mathcal{G}}+<ydg^{-1},\wedge dg>_{\mathcal{G}} 
\notag
\end{eqnarray}
for all $(q,p;g,y)\in T^{\ast }(M).$ Take now an element $\xi \in \mathcal{G}%
^{\ast }$ and assume that its isotropy subgroup $G_{\xi }=G,$ that is $%
Ad_{h}^{\ast }\xi =\xi $ for all $h\in G.$ In the general case such an
element $\xi \in \mathcal{G}^{\ast }$ can not exist but trivial $\xi =0,$ as
it happens to the Lie group $G=SL_{2}(\mathbb{R}).$ Then one can construct
the reduced phase space $l^{-1}(\xi )/G$ symplectomorphic to $(T^{\ast
}(N),\omega _{\xi }^{(2)}),$ where due to (0.12) for any $(q,p)\in T^{\ast
}(N)$%
\begin{eqnarray}
\omega _{\xi }^{(2)}(q,p) &=&<dp,\wedge dq>+<\Omega ^{(2)}(q),\xi >_{%
\mathcal{G}}  \TCItag{2.5} \\
&=&<dp,\wedge
dq>+\sum_{s=1}^{n}\sum_{i,j=1}^{3}e_{s}F_{ij}^{(s)}(q)dq^{i}\wedge dq^{j}. 
\notag
\end{eqnarray}
In the above we have expanded the element $\mathcal{G}^{\ast }\ni \xi
=\sum_{i=1}^{n}e_{i}a^{i}$ with respect to the bi-orthogonal basis $%
\{a^{i}\in \mathcal{G}^{\ast }:<a^{i},a_{j}>_{\mathcal{G}}=\delta _{j}^{i},$ 
$i,j=\overline{1,n}\}$ with $e_{i}\in \mathbb{R},$ $i=\overline{1,3},$ being
some constants, as well we denoted by $F_{ij}^{(s)}(q),$ $i,j=\overline{1,n}%
, $ $s=\overline{1,n},$ the corresponding curvature 2-form $\Omega ^{(2)}\in
\Lambda ^{2}(N)\otimes \mathcal{G}$ components, that is 
\begin{equation}
\Omega ^{(2)}(q):=\sum_{s=1}^{n}\sum_{i,j=1}^{3}a_{s\text{ }%
}F_{ij}^{(s)}(q)dq^{i}\wedge dq^{j}  \tag{2.6}
\end{equation}
for any point $q\in N.$ Summarizing calculations accomplished above, we can
formulate the following result.

\textbf{Theorem 2.1}\textit{\ Suppose a Yang-Mills field (2.1) on the fiber
bundle }$p:M\rightarrow N$\textit{\ with }$M=D\times G$\textit{\ is
invariant with respect to the Lie group }$G$\textit{\ action }$G\times
M\rightarrow M.$\textit{\ Suppose also that an element }$\xi \in G^{\ast }$%
\textit{\ is chosen so that }$Ad_{G}^{\ast }\xi =\xi .$\textit{\ Then for
the naturally constructed momentum mapping }$l:T^{\ast }(M)\rightarrow
G^{\ast }$\textit{\ (being equivariant) the reduced phase space }$l^{-1}(\xi
)/G\simeq T^{\ast }(N)$\textit{\ is endowed with the canonical symplectic
structure (2.5), having the following component-wise Poissoin brackets form:}

\begin{equation}
\{p_{i},q^{j}\}_{\xi }=\delta _{i}^{j},\text{ \ \ }\{q^{i},q^{j}\}_{\xi
}=0,\ \ \{p_{i},p_{j}\}_{\xi }=\sum_{s=1}^{n}e_{s}F_{ji}^{(s)}(q)  \tag{2.7}
\end{equation}
\textit{for all }$i,j=\overline{1,3}$\textit{\ and }$(q,p)\in T^{\ast }(N).$

The correspondingly extended Poisson bracket on the whole cotangent space $%
T^{\ast }(M)$ amounts due to (2.4) into the following set of Poisson
relationships: 
\begin{eqnarray}
\{y_{s},y_{k}\} &=&\sum_{r}^{n}c_{sk\text{ }}^{r}y_{r},\text{ \ \ \ \ \ \ \
\ \ \ \ \ }\ \ \{p_{i},q^{j}\}=\ \delta _{i}^{j}\ \ ,\text{ }  \TCItag{2.8}
\\
\text{\ }\{y_{s},p_{j}\} &=&0=\{q^{i},q^{j}\},\text{\ \ }\{p_{i},p_{j}\}=%
\sum_{s=1}^{n}y_{s\text{ }}F_{ji}^{(s)}(q),  \notag
\end{eqnarray}
where $i,j=\overline{1,3},$ $\ c_{sk}^{r}\in \mathbb{R},$ \ $s,k,r=\overline{%
1,n},$ are the structure constants of the Lie algebra $\mathcal{G}$, and we
made use of the expansion \ $A^{(s)}(q)=\sum_{j=1}^{3}A_{j}^{(s)}(q)$ $%
dq^{j} $ as well made changeable values $e_{i}:=y_{i},$ $i=\overline{1,n}.$
The result (2.8) can bee seen easily if one to rewrite the expression (2.4)
into an extended form as $\omega ^{(2)}:=\omega _{ext}^{(2)},$ where $\omega
_{ext}^{(2)}:=\left. \omega ^{(2)}\right| _{\mathcal{A}_{0}\rightarrow 
\mathcal{A}}$ $,$ $\mathcal{A}_{0}(g):=g^{-1}dg,$ $g\in G.$ Thereby one can
obtain in virtue of the invariance properties of the connection $\Gamma (%
\mathcal{A})$ that 
\begin{equation*}
\omega _{ext}^{(2)}(q,p;u,y)=<dp,\wedge dq>+d<y(g),Ad_{g^{-1}}\mathcal{A}%
(q;e)>_{\mathcal{G}}
\end{equation*}
\begin{equation*}
=<dp,\wedge dq>+<d\text{ }Ad_{g^{-1}}^{\ast }y(g),\wedge \mathcal{A}(q;e)>_{%
\mathcal{G}}=<dp,\wedge dq>+\sum_{s=1}^{n}dy_{s}\wedge du^{s}
\end{equation*}

\begin{equation*}
+\sum_{j=1}^{3}\sum_{s=1}^{nj}A_{j}^{(s)}(q)dy_{s}\wedge
dq-<Ad_{g^{-1}}^{\ast }y(g),\mathcal{A}(q,e)\wedge \mathcal{A}(q,e)>_{%
\mathcal{G}}
\end{equation*}
\begin{equation}
+\sum_{k\geq s=1}^{n}\sum_{l=1}^{n}y_{l}\text{ }c_{sk}^{l}\text{ }%
du^{k}\wedge du^{s}+\sum_{k=1}^{n}\sum_{i\geq
j=1}y_{s}F_{ij}^{(s)}(q)dq^{i}\wedge dq^{j},  \tag{2.9}
\end{equation}
where coordinate points $(q,p;u,y)\in T^{\ast }(M)$ \ are defined as
follows: $\mathcal{A}_{0}(e):=\sum_{s=1}^{n}du^{i}$ $a_{i},$ $%
Ad_{g^{-1}}^{\ast }y(g)=y(e):=\sum_{s=1}^{n}y_{s}$ $a^{s}$ for any element $%
g\in G.$ Whence one gets right away the Poisson brackets (2.8) plus
additional brackets connected with conjugated sets of variables $\{u^{s}\in 
\mathbb{R}:$\texttt{\ }$s=\overline{1,n}\}$ $\in \mathcal{G}$ $^{\ast }$\
and $\{y_{s}\in \mathbb{R}:$\texttt{\ }$s=\overline{1,n}\}\in \mathcal{G}:$

\begin{equation}
\{y_{s},u^{k}\}=\delta _{s}^{k},\text{ \ }\{u^{k},q^{j}\}=0,\text{ \ }%
\{p_{j},u^{s}\}=A_{j}^{(s)}(q),\text{ \ }\{u^{s},u^{k}\}=0,  \tag{2.10}
\end{equation}
where $j=\overline{1,3},$ \ $k,s=\overline{1,n},\ $and $\ \ q\in N.$

Note here that the suggested above transition from the symplectic structure $%
\omega ^{(2)}$ \ on $T^{\ast }(N)$ to its extension $\omega _{ext}^{(2)}$ on
\ $T^{\ast }(M)$ just consists formally in adding to the symplectic
structure $\omega ^{(2)}$ \ an exact part, which transforms it into
equivalent one. Looking now at the expressions (2.9) , one can infer
immediately that an element \ $\xi :=\sum_{s=1}^{n}e_{s}a^{s}\in \mathcal{G}%
^{\ast }$ \ will be invariant with respect to the $Ad^{\ast }$-action of the
Lie group $\ G$ \ iff 
\begin{equation}
\left. \{y_{s},y_{k}\}\right| _{y_{s}=e_{s}}=\sum_{r=1}^{n}c_{sk}^{r}\text{ }%
e_{r}\text{ }\equiv 0  \tag{2.11}
\end{equation}
identically for all $s,k=\overline{1,n},$ \ $j=\overline{1,3}$ \ and $\ q\in
N.$ In this and only this case the reduction scheme elaborated above will go
through.

Returning attention to the expression (2.10), one can easily write down the
following exact expression:

\begin{equation}
\omega _{ext}^{(2)}(q,p;u,y)=\omega ^{(2)}(q,p+\sum_{s=1}^{n}y_{s}\text{ }%
A^{(s)}(q)\text{ };u,y),  \tag{2.12}
\end{equation}
on the phase space $T^{\ast }(M)\ni (q,p;u,y),$ where we abbreviated for
brevity $<A^{(s)}(q),dq>$ as $\sum_{j=1}^{3}A_{j}^{(s)}(q)$ $dq^{j}.$ The
transformation like (2.12) \ was discussed within somewhat different context
in article [8] containing also a good background for the infinite
dimensional generalization of symplectic structure techniques. Having
observed from (2.12) that the simple change of variable 
\begin{equation}
\tilde{p}:=p+\sum_{s=1}^{n}y_{s}\text{ }A^{(s)}(q)  \tag{2.13}
\end{equation}
of the cotangent space $T^{\ast }(N)$ recasts our symplectic structure (2.9)
into the old canonical form (2.4), one obtains that the following new set of
Poisson brackets on $T^{\ast }(M)$ $\ni (q,\tilde{p};u,y):$%
\begin{eqnarray}
\{y_{s},y_{k}\} &=&\sum_{r=1}^{n}c_{sk}^{r}\text{ }y_{r},\text{ \ \ \ }\{%
\tilde{p}_{i},\tilde{p}_{j}\}=0,\text{\ \ \ \ \ }\{\tilde{p}%
_{i},q^{j}\}=\delta _{i}^{j},\text{ }  \TCItag{2.14} \\
\{y_{s},q^{j}\} &=&0\text{ }=\{\tilde{p}_{i},\tilde{p}_{j}\},\text{\ }%
\{u^{s},u^{k}\}=0,\text{ \ \ }\{y_{s},\tilde{p}_{j}\}=0,\text{ \ }  \notag \\
\{y_{s},q^{i}\} &=&0,\text{ \ \ \ \ \ \ \ \ \ }\{y_{s},u^{k}\}=\delta
_{s}^{k},\text{ \ \ \ \ \ \ \ \ \ \ }\{u^{s},\tilde{p}_{j}\}=0,  \notag
\end{eqnarray}
where $\ k,s=\overline{1,n}$ \ and $i,j=\overline{1,3}$, holds iff the
Yang-Mills equations 
\begin{equation}
\partial F_{ij}^{(s)}/\partial q^{l}+\partial F_{jl}^{(s)}/\partial
q^{i}+\partial F_{li}^{(s)}/\partial q^{j}  \tag{2.15}
\end{equation}
\begin{equation*}
+\sum_{k,r=1}^{n}c_{kr}^{s}(F_{ij}^{(k)}A_{l}^{(r)}+F_{jl}^{(k)}A_{i}^{(r)}+F_{li}^{(k)}A_{j}^{(r)})=0
\end{equation*}
are fulfilled for all $\ s=\overline{1,n}$ \ and  $i,j,l=\overline{1,3}$ on
the base manifold $\ N.$ This effect of complete reduction of Yang-Mills
variables from the symplectic structure (2.9) is known in literature [1,8]
as the principle of minimal interaction and appeared to be useful enough for
studying different interacting systems as in [9,10]. In part 2 of this work
we shall continue a study of reduced symplectic structures connected with
infinite dimensional coupled dynamical systems like Yang--Mills -Vlasov,
Yang-Mills- Bogoliubov and Yang-Mills-Josephson ones.

\textbf{3.Acknowledgements.}

One of the authors (A.P.) is cordially indebted to Prof. B.A. Kupershmidt
for sending a set of inspiring reprints of his articles some of which are
cited through this work, and as well to Prof. J.\ Zagrodzinski for sending
his article before publication. Special thanks for nice hospitality and warm
research atmosphere are due to the staff of Dept. of Physics at the EMU of
N.Cyprus, especially to Profs. M. Halilsoy and A. Istillozlu for valuable
discussions of problems under regard.

\end{document}